\documentclass[superscriptaddress,showpacs,prl,twocolumn,floatfix,twoside,tbtags]{revtex4}

\usepackage[utf8]{inputenc}
\usepackage[T1]{fontenc}
\usepackage{subfigure}
\usepackage{graphicx}
\usepackage{amsmath,amssymb}
\usepackage{xspace}
\DeclareMathOperator{\grandO}{O}

\newcommand{\etal}{\textit{et al}.\@\xspace}
\newcommand{\ie}{\textit{i.e.}\@\xspace}

\begin{document}

\title{Dislocation core energies and core fields from first principles}

\author{Emmanuel \surname{Clouet}}
\affiliation{CEA, DEN, Service de Recherches de Métallurgie Physique, 
F-91191 Gif-sur-Yvette, France}
\affiliation{Laboratoire de Métallurgie Physique et Génie des Matériaux,
UMR CNRS 8517, Université de Lille 1, 59655 Villeneuve d'Ascq, France}

\author{Lisa \surname{Ventelon}}
\affiliation{CEA, DEN, Service de Recherches de Métallurgie Physique, 
F-91191 Gif-sur-Yvette, France}

\author{F. \surname{Willaime}}
\affiliation{CEA, DEN, Service de Recherches de Métallurgie Physique, 
F-91191 Gif-sur-Yvette, France}


\date{\today}
\begin{abstract}
Ab initio calculations in bcc iron show that a $\left<111\right>$ screw dislocation induces a short range dilatation field in addition to the Volterra elastic field. This core field is modeled in anisotropic elastic theory using force dipoles. The elastic modeling thus better reproduces the atom displacements observed in ab initio calculations. Including this core field in the computation of the elastic energy allows deriving a core energy which converges faster with the cell size, thus leading to a result which does not depend on the geometry of the dislocation array used for the simulation.
\end{abstract}
\maketitle

Plastic deformation in crystals is heavily related to the dislocation core properties \cite{HIR82}.
As experimental investigation of the dislocation core is difficult,
atomic simulations have become a common tool in dislocation theory. 
But dislocations induce a long-range elastic field and
one has to take full account of it in the atomic modeling.
This is even more crucial for ab initio calculations 
because of the small size of the unit cell that can be simulated.
In this Letter, we illustrate this point for the screw dislocation in bcc Fe
by showing that the commonly-used elastic description, 
\ie the Volterra solution \cite{HIR82}, 
has to be enriched in order to get quantitative information
from ab initio calculations. 

Two different methods based on ab initio calculations
have been developed to model dislocations. 
In the first approach,
a single dislocation is introduced in a unit cell 
which is periodic only along the dislocation line
and with surfaces in the other directions. 
Surface atoms are displaced according to the dislocation long-range
elastic field and can be either kept fixed
or relaxed using lattice Green functions \cite{WOO01b}. 
The main drawback of this method is that,
in ab initio calculations, one cannot separate 
the energy contribution of the dislocation from the surface one.
To calculate dislocation energy properties,
one has to use the second approach which is based on full periodic 
boundary conditions \cite{ISM00,FRE03,BLA00,CAI01}.
As this is possible only if the total Burgers vector of the unit cell is zero,
a dislocation dipole is simulated.
Using elasticity theory, one can calculate the interaction
between the two dislocations forming the dipole 
as well as with their periodic images \cite{CAI01},
and thus isolate dislocation intrinsic properties.

We use this dipole approach to study the core properties 
of $\left<111\right>$ screw dislocations in bcc iron 
with ab initio calculations based on density functional theory 
using the SIESTA code as described in Ref.~\onlinecite{VEN07}.
The dislocations are positioned
at the center of gravity of three neighboring atomic columns.
Depending on the sign of the Burgers vector compared to the helicity of the original site,
there are two different configurations, termed ``easy'' and ``hard''.
The ``hard'' core configuration shifts locally the atoms 
such that they lie in the same $\left\{111\right\}$ plane.
From steric considerations, one thus expects this configuration to be 
less stable.
The energy landscape experienced by the gliding dislocation
is dictated by the energy difference between these two configurations
which is
a maximum for the Peierls barrier.
It is therefore important to get a precise knowledge of the corresponding core energies.

\begin{figure}[bt]
	\begin{center}
		{\includegraphics[height=0.99\linewidth,angle=-90]{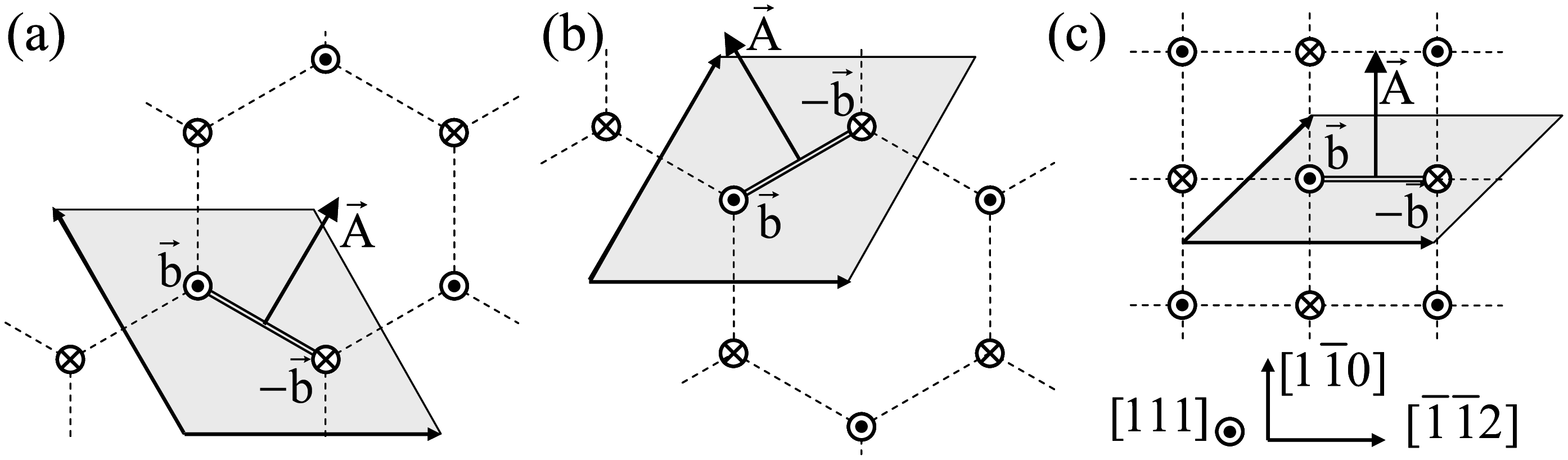}}
	\end{center}
	\caption{Screw dislocation periodic arrangements used for ab initio calculations:
	(a) T and (b) AT triangular arrangements;
	(c) quadrupolar arrangement. 
	$\vec{b}=\frac{1}{2}\left[ 111 \right]$ for 
	``easy'' and $\frac{1}{2}\left[ \bar{1}\bar{1}\bar{1} \right]$ 
	for ``hard'' cores.
	$\vec{A}$ is the dipole cut vector.}
	\label{fig:PBC}
\end{figure}

We introduce the dipole in periodic unit cells
corresponding to different dislocation arrays \cite{VEN07}.
The triangular arrangements of Figs.~\ref{fig:PBC}a and \ref{fig:PBC}b
preserve the 3-fold symmetry of the bcc lattice in the $\left[111\right]$ direction.
One can obtain two variants which are related by a $\pi/3$ rotation.
We refer to them as the twinning (T) (Fig.~\ref{fig:PBC}a)
and the anti-twinning (AT) triangular arrangement (Fig.~\ref{fig:PBC}b)
\cite{EndNote2}.
The last dislocation arrangement (Fig.~\ref{fig:PBC}c)
is equivalent to a rectangular array of quadrupoles.

\begin{figure}[bt]
	\begin{center}
		\includegraphics[width=0.95\linewidth]{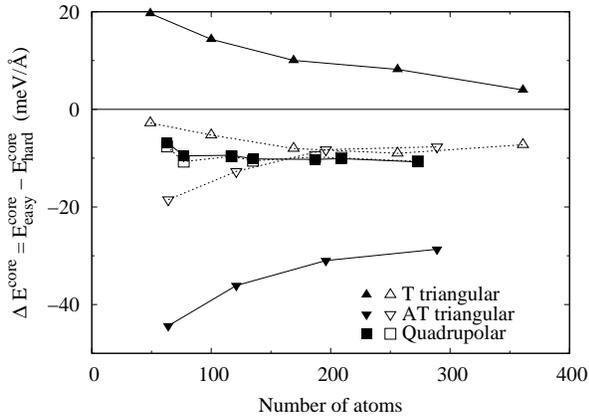}
	\end{center}
	\caption{Core energy difference between the ``easy'' and ``hard''
		core configurations.
		Solid symbols correspond to core energies obtained when only 
		the Volterra field is considered 
		and open symbols to core energies when both the Volterra and
		the core fields are taken into account ($r_{c}=3$~\AA).}
	\label{fig:Ecore}
\end{figure}

Simulation unit cells are built so that the two dislocations composing the dipole
are in the same configuration, either ``easy'' or ``hard'' depending on
the sign of the Burgers vector.
Assuming that the elastic displacement field created by each dislocation
corresponds to the Volterra one \cite{HIR82}, the elastic
energy stored in the simulation box is proportional to the square
of the Burgers vector and therefore is the same for 
the ``easy'' and ``hard'' configurations.
The core energy difference between the two possible configurations
is thus simply given by half the energy difference obtained from ab initio 
calculations for the same unit cell.
This energy difference is shown as solid symbols in Fig.~\ref{fig:Ecore}. 
The result depends on the chosen dislocation arrangement.
According to the T triangular arrangement, the ``hard''
core configuration is more stable than the ``easy'' one, 
whereas the quadrupolar and the AT triangular arrangements
lead to the opposite conclusion. 
For a given arrangement, the convergence with the number $N$ of atoms
is proportional to $N^{-1/2}$. 
The computational cost to directly deduce converged values 
from ab initio calculations is therefore out of reach.

\begin{figure*}[!t]
	\begin{center}
		\includegraphics[width=0.99\linewidth]{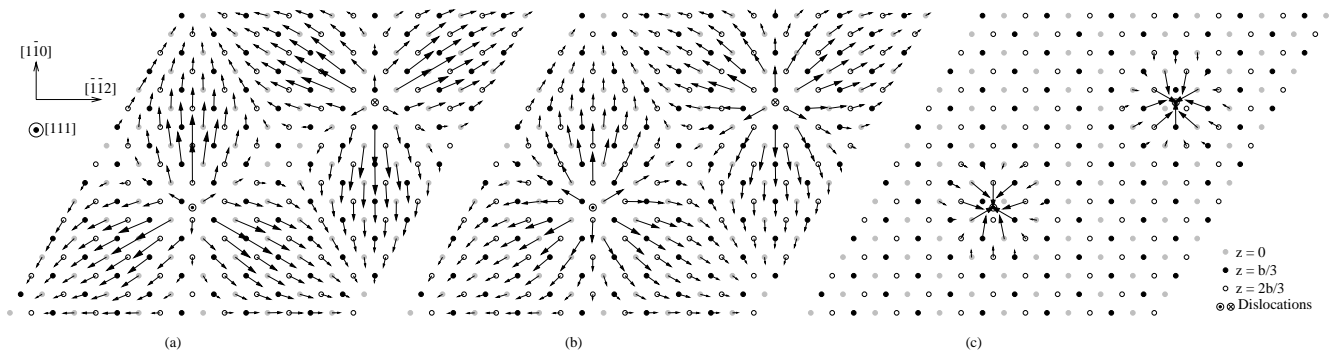}
	\end{center}
	\caption{Planar displacement map of a periodic unit cell 
	containing a screw dislocation dipole obtained from ab initio calculations:
	(a) total displacement, 
	(b) after subtraction of the Volterra elastic field,
	and (c) after subtraction of the Volterra and the core elastic fields.
	Vectors correspond to $(111)$ in-plane displacements and have been magnified by a factor 50.
	Displacements smaller than 0.01~{\AA} are omitted. 
	For clarity, displacements of the six atoms belonging to the cores of the two dislocations 
	are not shown in (c). 
	Atomic positions are drawn as circles with a color depending of their original $(111)$ plane.}
	\label{fig:map}
\end{figure*}

To understand how our simulation approach has to be enriched 
to lead to unambiguous dislocation core energies,
we examine the atom displacements created by the dislocation array 
in ab initio calculations.
For all unit cells, atom displacements in the $[111]$ direction,
\ie the screw component, correspond to dislocations having a symmetrical 
and compact core structure,
in agreement with recent ab initio calculations
in bcc Fe \cite{FRE03,DOM05,VEN07}.
The screw dislocation dipoles also create displacements in the $(111)$ plane, 
\ie perpendicular to the screw axis (Fig.~\ref{fig:map}a). 
Part of this edge component arises from elastic anisotropy. 
Nevertheless, when subtracting the displacements
predicted by anisotropic elasticity for the periodic dislocation array \cite{CAI01}
from the ones given by ab initio calculations,
one obtains a residual displacement (Fig.~\ref{fig:map}b) 
which looks like a combination 
of 2-dimension expansions centered at the dislocations.

This is not included in the Volterra solution 
describing the dislocation elastic field.
Nevertheless, going back to the seminal paper of Eshelby \etal \cite{ESH53},
it appears that a dislocation can also lead to such a supplementary elastic field.
Indeed, Eshelby \etal showed that 
a straight dislocation in an infinite elastic medium 
creates in a point defined by its cylindrical coordinates $r$ and $\theta$
a displacement given 
by a Laurent series which leading terms are
\begin{equation}
	\vec{u}(r,\theta) = 
	\vec{v}\ln(r) 
	+ \vec{u}_0(\theta)
	+ \vec{u}_1(\theta)\frac{1}{r}
	+ \grandO{\left(\frac{1}{r^2}\right)}.
	\label{eq:displacement}
\end{equation}
Usually, only the two first terms of this series are considered 
leading to the well-known Volterra solution \cite{STR58}. 
This gives the long-range displacement induced 
by the discontinuity along the dislocation cut.

Close to the dislocation core, the third term in Eq.~\ref{eq:displacement}
may be relevant too \cite{GEH72,HEN05}.
This corresponds to what is usually called the dislocation core field.
Such a field arises 
from non-linearities in the crystal elastic behavior
and from perturbations due to the atomic nature of the core.
It can be modeled within anisotropic linear elasticity theory
using line-force dipoles representative 
of an elliptical line source expansion
located close to the dislocation core \cite{HIR73}.
The core field is then characterized by 
the first moments $M_{ij}$ of this line-force distribution.
We propose in the following an original  approach that allows
to directly deduce the moments $M_{ij}$ from quantities that can be 
``measured'' in atomic simulations.

In that purpose, we consider the elastic energy of a periodic unit cell containing a dislocation dipole
defined by its Burgers vector $\vec{b}$ and its cut vector $\vec{A}$.
Each dislocation also creates a core field corresponding to the moments 
given by the second-rank tensor $M$.
An homogeneous strain can be superposed to the heterogeneous 
strain created by the dislocation dipole.
This contributes to the elastic energy by an amount \cite{BAC80}
\begin{equation}
	E_{\varepsilon} = h\left(  \frac{1}{2} S C_{ijkl}\varepsilon_{ij}\varepsilon_{kl}
	+ C_{ijkl}b_{i}A_{j}\varepsilon_{kl}
	- 2 M_{ij}\varepsilon_{ij} \right) ,
	\label{eq:Estrain}
\end{equation}
where $S$ is the area of the simulation unit cell perpendicular to the dislocation lines,
$h$ the corresponding height 
and $C_{ijkl}$ the elastic constants.
The homogeneous stress is defined as
\begin{equation}
	\sigma_{ij} = \frac{1}{hS}\frac{\partial{E_{\varepsilon}}}{\partial{\varepsilon_{ij}}}
	= C_{ijkl}(\varepsilon_{kl}-\varepsilon_{kl}^0) ,
	\label{eq:sigma}
\end{equation}
with the stress-free strain 
\begin{equation}
	\varepsilon_{ij}^0 = - \frac{b_i A_j + b_j A_i}{2S} + 2 S_{ijkl} \frac{M_{kl}}{S},
	\label{eq:epsi0}
\end{equation}
where the elastic compliances $S_{ijkl}$ are the inverse of the elastic constants.

When the dislocations do not create any core field ($M=0$), 
one recovers the fact that  the elastic energy is minimal
for an homogeneous strain equal to the plastic strain 
produced when the dislocation 
dipole is introduced in the simulation unit cell \cite{CAI01}. 
The core fields induce a second contribution 
which is proportional to the dislocation density,
thus allowing to define a dislocation formation volume.
Our ab initio calculations lead for a screw dislocation in bcc iron 
to a dilatation perpendicular to the dislocation line, 
$\delta V_{\bot}=(\varepsilon_{11}^0+\varepsilon_{22}^0)S/2
=3.8\pm0.3$~\AA$^2$,
and to a contraction along the dislocation line,
$\delta V_{//}=\varepsilon_{33}^0S/2=-1.3\pm0.2$~\AA$^2$,
where the formation volumes are defined per unit of dislocation line.

Instead of letting the unit cell relax its size and shape,
one can also keep fixed the periodicity vectors and minimize
the energy only with respect to the atomic positions.
The simulation box is thus subject to an homogeneous stress
from which the moments responsible for the dislocation core field
can be deduced using Eqs.~\ref{eq:sigma} and \ref{eq:epsi0}.
The component $\sigma_{33}$ of the ``measured'' homogeneous stress 
is negligible compared to 
$\sigma_{11}$ and $\sigma_{22}$, in agreement with the following argument.

For a $\left[111\right]$ screw dislocation in a cubic crystal, 
because of the 3-fold symmetry, the tensor $M$ is diagonal 
with $M_{11}=M_{22}$ and $M_{33}=0$ if the unit vector $\vec{e}_3$ corresponds 
to the $\left[111\right]$ direction.
The core field is thus a pure dilatation in the $\left(111\right)$ plane. 
This is true when the dislocation is in a stress-free state 
or if the stress experienced by the dislocation also obeys this 3-fold symmetry.
The ab initio calculations indeed lead to such a 
tensor $M$ for the two triangular arrangements.
The quadrupolar arrangement induces a stress
which does not obey this symmetry. Because of the moment polarizability \cite{EndNote3}, 
we obtain different values for $M_{11}$ and $M_{22}$ in this case.
Nevertheless, all dislocation arrangements used in ab initio calculations
converge with the cell size 
to $M_{11}=M_{22}=650\pm50$~GPa.\AA$^2$
for both ``easy'' and ``hard'' core configurations.
As for the contraction observed along the dislocation line,
it arises from the elastic compliance $S_{1133}$
which couples the strain component $\varepsilon_{33}$
with the force moments $M_{11}$ and $M_{22}$.

Knowing the moments, we model the dislocation elastic displacement
as the superposition of the Volterra and the core fields.
We can thus compare the displacement given by ab initio calculations
with the one predicted by elasticity theory for the dislocation periodic array \cite{CAI01}.
Looking at the difference between the fields given by the two modeling techniques
for the in-plane $(111)$ component (Fig.~\ref{fig:map}c), 
one sees that elasticity theory perfectly manages to reproduce the displacement given 
by ab initio calculations, except for atoms which are too close to the dislocation cores.
It is clear that the superposition of the core field to the Volterra solution
greatly improves the description of the dislocation elastic field.

The excess energy $E$, \ie the energy difference per unit of height
between the unit cell with and without the dislocation dipole,
is the sum of the two dislocation core energies $E^{\mathrm{core}}$
and of the elastic energy. 
\begin{equation}
	E = 2E^{\mathrm{core}} 
	+ E^0 - b_i K^0_{ij} b_j \ln{(r_c)}
	+ M_{ij} K^2_{ijkl} M_{kl} \frac{1}{{r_c}^2}
	\label{eq:energy}
\end{equation}
where $K^0$ and $K^2$ are definite positive tensors 
which only depend on the elastic constants.
$E^0$ 
contains 
the elastic interaction
between the two dislocations composing the primary dipole, as well 
as the interaction with their periodic images.
The core fields modify this interaction energy
as the dislocations now interact not only through their Volterra elastic fields
but also through their core fields and the combination of these two elastic fields. 
The last term in Eq.~\ref{eq:energy} corresponds to the increase 
of the dislocation self elastic energy due to their core fields.
The cutoff distance $r_c$ is introduced because elastic fields
are diverging due to elasticity inability to describe atom displacements
in the dislocation core.
The core energy that can be deduced from atomic simulations therefore 
depends on the value of $r_c$. 

\begin{figure}[bt]
	\begin{center}
		\includegraphics[height=0.90\linewidth,angle=-90]{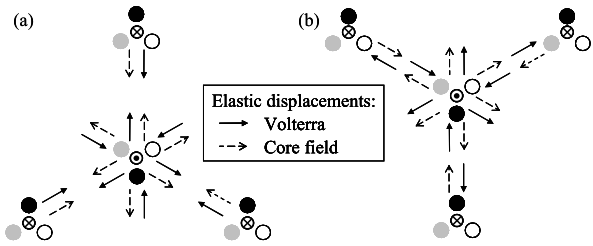}
	\end{center}
	\caption{Sketch of the $(111)$ in-plane displacement created
	by the triangular arrangement of dislocations in their ``easy''
	core configuration.
	For the T variant (a), the displacements due to 
	the  Volterra and the core fields have the same sign 
	and sum up in the region between two neighboring dislocations,
	whereas they partially cancel for the AT variant (b).}
	\label{fig:interaction}
\end{figure}

We use Eq.~\ref{eq:energy} to extract dislocation core energy
from atomic simulations: $E$ and $M$ are deduced from ab initio calculations,
whereas $E^0$, $K^0$, and $K^2$ are calculated with anisotropic elasticity theory.
A core radius  slightly larger than the Burgers vector ($r_c=3$~\AA)
leads to reasonable core energies 
and a good convergence with the size of the simulation unit cell.
The core energy difference between the ``easy'' and the ``hard'' core configurations
of the screw dislocation in bcc iron 
converges now rapidly to a value which does not depend on the geometry of the dislocation 
arrangement (Fig.~\ref{fig:Ecore}).
For all simulations, the ``easy'' core configuration is more stable than
the ``hard'' one, with a core energy converging respectively to 
$E_{\mathrm{easy}}^{\mathrm{core}}=219\pm1$~meV.\AA$^{-1}$
and $E_{\mathrm{hard}}^{\mathrm{core}}=227\pm1$~meV.\AA$^{-1}$.

We can now understand why the simple approach, 
where only the Volterra elastic field is considered,
leads to core energies which strongly depend on the geometry 
of the dislocation array. 
Looking at the $(111)$ in-plane displacement created by each component of the dislocation elastic field,
the Volterra part oscillates as a function of $\theta$ 
between a compression and a tension type, 
whereas the core-field only leads to a compression.
This is illustrated in Fig.~\ref{fig:interaction} for the two variants 
of the triangular arrangement with the dislocations in their ``easy'' core 
configuration. It is clear on this figure, that the effects of the Volterra and the core fields 
will sum up in the regions between two neighboring dislocations for 
the T variant (Fig.~\ref{fig:interaction}a), whereas they will partially 
compensate for the AT variant (Fig.~\ref{fig:interaction}b).
One thus expects a stronger elastic interaction between dislocations for the T 
variant than for the AT one.
This is the opposite for the ``hard'' core configuration, as changing the sign
of the Burgers vector reverses the Volterra elastic field without modifying
the core field.
When neglecting the dislocation core field, one thus overestimates the elastic 
energy difference between the ``easy'' and ``hard'' core configurations for the 
T variant and underestimates it for the AT one.
On the other hand, the coupling of the Volterra and the core elastic field 
leads to a negligible interaction between neighboring dislocations for the 
quadrupolar arrangement because of its centro-symmetry. 
This arrangement actually appears as the best-suited one
to extract quantitative information from atomic simulations \cite{CAI01}.

This dilatation due to the dislocation core field is not specific to iron.
When analyzing previous ab initio calculations \cite{ISM00,FRE03}
we can conclude that screw dislocations exhibit 
a similar core field in other bcc metals like Mo and Ta.
On the other hand, empirical potentials may fail to predict such a core dilatation.
This is the case for Mendelev potential \cite{MEN03} 
which is often used to study dislocations in iron
\cite{DOM05,CHA06,CLO08}.

In addition to determining the formation volume of the screw dislocation in iron
and modeling it within anisotropic linear elasticity theory,
our study shows that considering this core field is crucial
when deriving from atomic simulations dislocation parameters
like their core energy.
This supplementary elastic field should also influence
any energy differences, like the Peierls barriers,
and stresses extracted from atomic simulations. 
Moreover, because of the formation volume associated with this core field,
a dislocation can interact with an hydrostatic stress. 
Close to the core, it will modify the dislocation interaction with point defects \cite{CLO08}.

\begin{acknowledgments}
	This work was supported by the European Fusion Materials Modeling
	program and by the SIMDIM project under contract ANR-06-BLAN-250.
\end{acknowledgments}


\begin{thebibliography}{20}
\bibitem{EndNote3}
	The dipole polarizability describes its moment variation
	with the locally applied strain.
\end{thebibliography}
\end{document}